\documentclass[prc,twocolumn,secnumroman,tightenlines,amssymb,aps,nobibnotes,superscriptaddress,balancelastpage,floatfix,nofootinbib]{revtex4}
\usepackage{soul}  
\usepackage{parskip}
\usepackage{amsmath}
\usepackage{slashbox}
\usepackage{soul}
\usepackage{float}
\usepackage{lipsum}
\usepackage{graphicx}
\usepackage{multirow}           
\usepackage{color}
\DeclareRobustCommand\substyle{\name@idx{document substyle}}
\DeclareRobustCommand\classoption{\name@idx{document class option}}
\DeclareRobustCommand\classname{\name@idx{document class}}
\def\name@idx#1#2{{\ttfamily#2}
\index{#2\space#1=\string\ttt{#2}\space#1}\index{#1>#2=\string\ttt{#2}}}

\DeclareTextFontCommand{\rb}{\color{red}\bfseries}

\begin{document}







\title{Neutron emission associated with $\gamma\gamma\to\gamma\gamma$ scattering in UPC of heavy ions at the LHC}

\author{P. Jucha}
\email{Pawel.Jucha@ifj.edu.pl}
\affiliation{Institute of Nuclear Physics PAN, ul.\,Radzikowskiego 152,PL-31-342 Krak\'ow, Poland}

\author{M. K\l{}usek-Gawenda}
\email{Mariola.Klusek@ifj.edu.pl }
\affiliation{Institute of Nuclear Physics PAN, ul.\,Radzikowskiego 152,PL-31-342 Krak\'ow, Poland}

\author{A. Szczurek}
\email{Antoni.Szczurek@ifj.edu.pl}
\affiliation{Institute of Nuclear Physics PAN, ul.\,Radzikowskiego 152,PL-31-342 Krak\'ow, Poland}
\affiliation{Institute of Physics, Faculty of Exact and Technical Sciences, University of Rzesz\'ow, ul. Pigonia 1, PL-35-310 Rzesz\'ow, Poland}

\date{\today}

\begin{abstract}
We calculate cross sections for the production of different neutron
classes $0n 0n$, $0nXn+Xn0n$ and $Xn Xn$, possible to
measure in Zero Degree Calorimeters (ZDCs), associated with photon-photon scattering in UPC 
of lead on lead.
The calculations are performed for the ATLAS kinematics.
Our calculations for neutron classes are tested against existing
data for $\rho^0$ production in UPC. We get a good agreement for absolute
cross section as well as very good results for fractional 
cross section which gives confidence to the analogous calculation
for diphoton production, corresponding to 
$\gamma \gamma \to \gamma \gamma$ scattering.
We present both absolute as well as fractional cross sections
for the different neutron classes. 
We also show different distributions in impact parameter, photon rapidity, transverse momentum, 
diphoton invariant mass and photon rapidity difference. The shapes of the
photon distributions depend on the neutron classes which we quantify in
this paper. It would be valuable to test our predictions using the ATLAS main
detector and the ATLAS ZDCs.
\end{abstract}

\maketitle


\section{Introduction}
\label{Sect.I}
The light-by-light scattering (LbL) is a quantal
process which was ``measured'' only recently in ultraperipheral
collisions (UPC) of heavy ions at the LHC \cite{ATLAS1, CMS1, ATLAS2, ATLAS3, CMS_2024}.
Different mechanisms were considered by us in recent years
\cite{Klusek_Gawenda_2016,Nasza3_2gluo,LP}. The box mechanism was proven to be
the dominant mechanism for the ATLAS and CMS kinematics.
Therefore in the following we shall include only the box mechanism
(with leptons and quarks) formed in a fusion of two quasi-real photons.
 
The photon-photon scattering in UPC can be also studied at lower diphoton
masses with ALICE and LHCb detectors \cite{KSS} and in a somewhat
remote future with FoCal and ALICE 3 detectors \cite{FoCAL, ALICE3, Jucha2024}.
Taking into account the detector's acceptance, compared to the ATLAS and CMS
experiments, there may be other mechanisms.

Recently we also calculated inelastic processes where photon(s)
couple to individual nucleons (proton or neutron) \cite{KGS_2025}.
In such processes, called by us inelastic, one or even two nuclei are
excited and may emit probably a small number of neutrons.
Such processes constitute a background to purely coherent processes
when photons couple rather to nuclei.

In the present paper, we consider only coherent processes.
In this case emission of neutrons is caused by extra photon
exchanges which lead to excitation of one or both colliding
nuclei. The excited nuclei typically emit one or more neutrons
which are emitted in forward and backward directions due
to high-energy boosts. The formalism of neutron emission in UPC was
discussed in recent years  \cite{Baltz_2002,Baltz_2009,reldis,Klusek-Gawenda:2013ema, starlight, NooN, Strikman_2023, Jucha2025}.
Calculation of the emission of a given number of neutrons is more tricky and requires
modeling of the underlying nuclear processes. However, calculating the cross section for
a given class of neutron emissions is simpler and more reliable.
This will be discussed in the next section.

\section{Formalism}

The cross section for LbL associated with neutron emission in lead-lead UPC is calculated in
equivalent photon approximation in the impact parameter space as:
\begin{equation}
    \begin{split}
        \sigma_{PbPb\to AA+\gamma\gamma+kn+jn} = \int \frac{d\sigma_{\gamma\gamma\to\gamma\gamma}(W_{\gamma\gamma})}{dz}  S^2(b) P^{c}(b) \\ \times N(\omega_1,b_1)N(\omega_2,b_2)   
     d^2b d\bar{b}_x d\bar{b}_y \frac{W_{\gamma\gamma}}{2} dW_{\gamma\gamma}dY_{\gamma\gamma} dz \;.
\label{EPA_inbspace}
\end{split}
\end{equation}
In the outgoing channel, $A$ stands for the lead nucleus minus $k$ or $n$ neutrons. 
The impact parameter approach is very useful as here the probability
$P^c(b)$ of a given neutron category \textit{c} can be easily included by inserting it under
the integral in Eq. (\ref{EPA_inbspace}). $S^2(b)$ is a probability that is less than $1$
for $b < R_{Pb_1} + R_{Pb_2}$, and equals 1 for the $b > R_{Pb_1} + R_{Pb_2}$.  The survival probability for the rapidity gap can
be calculated as 

\begin{equation}
    S^2(b) = \text{exp}\big( -\sigma_{NN} \cdot T_{AB}(b) \big).
\end{equation}

\noindent where $\sigma_{NN}$ is the inelastic nucleon-nucleon cross section. The $T_{AB}(b)$ nuclear overlap function above means the convolution integral

\begin{equation}
   T_{AB}(b) = \int d^2s T_A(\textbf{s}) T_B(\textbf{b}-\textbf{s}),
\end{equation}

\noindent where $T_A$ and $T_B$ are standard nuclear matter thickness functions $T_A(\textbf{b}) = \int \rho(\vec{r}) dz$, where $\vec{r}=(\vec{b},z)$. 


The main goal of this research is to determine the contribution of the percentage
of final state ($\gamma\gamma$) production associated with the emission or absence of emission for a given process. For this purpose, we introduce a probability
of the neutron category $P^c(b)$. 
The theoretical model on the production of particle accompanied by
neutron emission was discussed e.g. in \cite{Baltz_2002, Baltz_2009}.
 We define the average number of photon interactions with a nucleus as

\begin{equation}
    m(b) = \int d\omega N(\omega,b) \sigma_{abs}(\omega),
\end{equation}

\noindent where $\sigma_{abs}(\omega)$ is energy dependent cross section for photon absorption.
The calculations use a parametrization that takes into account the description of the
physical sub-processes that give input to the photoabsorption cross section, including GDR,
quasi-deuteron, $\Delta$ resonance, see Ref. \cite{Klusek-Gawenda:2013ema}.
Now we can define quantities that form the basis of probability calculations. We assume that the sum of probabilities for the emission of at least one neutron $Xn$, and the absence of neutron $0n$, gives 1:

\begin{equation}
    \begin{split}
    &P^{Xn}(b) = 1-\text{exp}[-m(b)], \\
    &P^{0n}(b) = \text{exp}[-m(b)].
    \end{split}
\end{equation}

Having fixed the auxiliary probabilities we can calculate probabilities for
well-defined neutron categories: $0n0n$, $0nXn$, $Xn0n$, $XnXn$. The impact parameter
dependent probabilities can be easily calculated in an almost model-independent
way as:

\begin{equation}
    \begin{split}
    &P^{XnXn}(b) = \big(P^{Xn}(b) \big)^2, \\
    &P^{0n0n}(b) = \big(P^{0n}(b) \big)^2, \\
    &P^{Xn0n}(b) = P^{Xn}(b)P^{0n}(b). \\
    \end{split}
\end{equation}

The sum of the cross sections that takes into account all configurations of neutron emissions
gives the total cross section.


\section{Results}

\subsection{Break-up probabilities}

\begin{figure}
    \centering
    \includegraphics[width=0.46\textwidth]{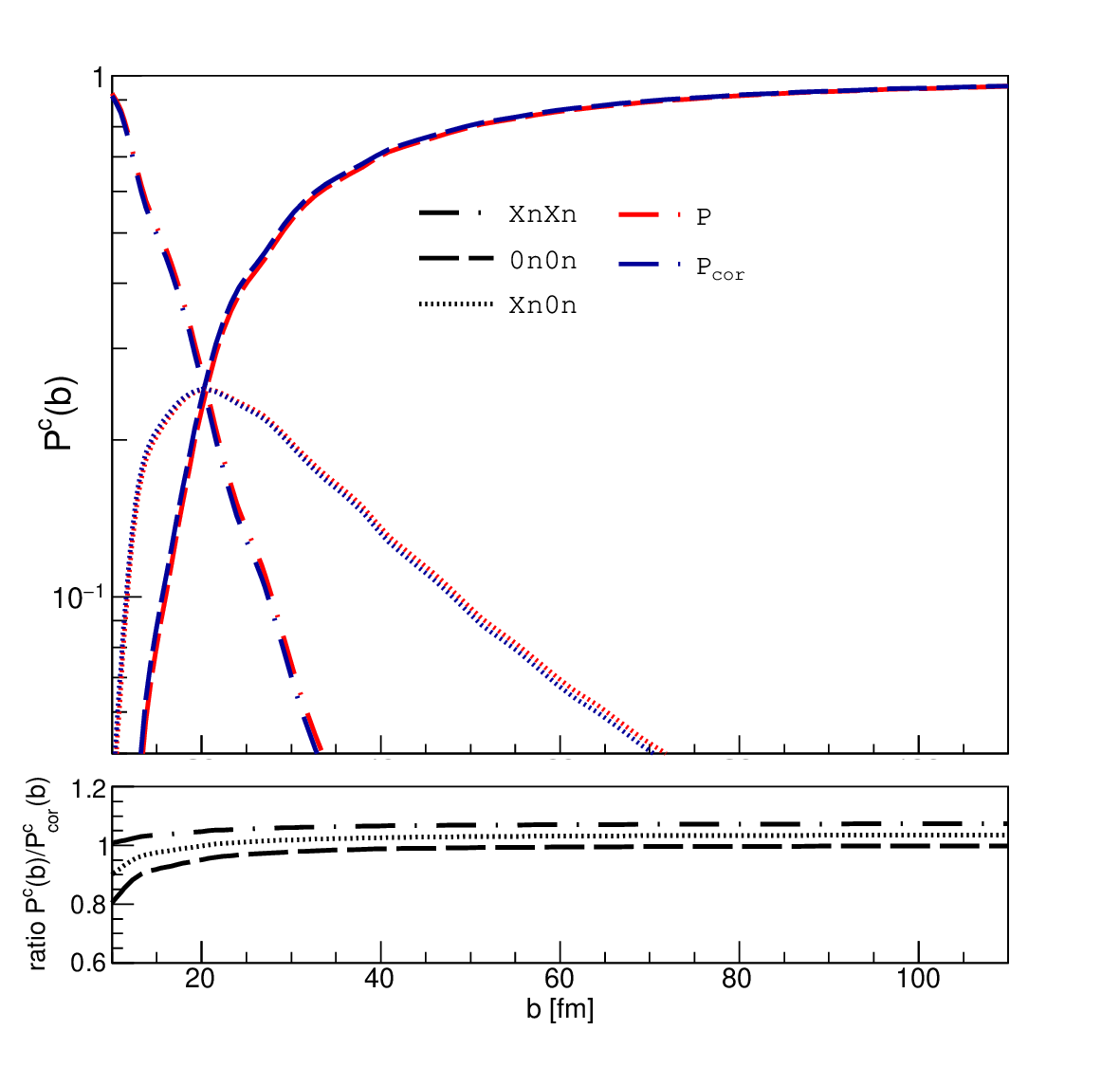}
    \caption{Break-up probabilities as a function of impact parameter:
      $XnXn$ (dash-dotted line), $0n0n$ (dashed line) and $Xn0n$ (dotted line). The lower panel is explained in the main text. }
\label{fig:probabilities}
\end{figure}

In Fig.\ref{fig:probabilities} we show probabilities defined
in the previous section as a function of impact parameter for all 
neutron categories. The probabilities are normalized
to fulfill the following relation:
\begin{equation}
P^{0n0n}(b) + P^{0nXn}(b) + P^{Xn0n}(b) + P^{XnXn}(b) = 1.
\end{equation}
At larger $b$ the $0n0n$ category clearly dominates, the $Xn0n$ category reaches its maximum
at $b \approx$ 20 fm. In contrast, at small impact parameters $\sim 2 R_A$
the $XnXn$ category is the primary one. One may say that selecting a category
allows to choose a characteristic impact parameter.
The results presented so far are based on the assumption that
absorption of photon necessarily leads to the emission of one or more
neutrons. However at low excitation energies also photons can
be emitted and compete with neutron emission from the excited nucleus. How important can be such
an effect we know only theoretically based on statistical model codes.
Recently the Experimental Storage Ring (ESR) group at GSI Darmstadt measured
the competition between photon and neutron emissions. The probability
of neutron emission was measured up to excitation energies 
$E_{exc}$ = 9.5 MeV \cite{GSI}. The probability increases 
from the threshold (neutron separation energy) and slowly
converges to 1. We have parametrized the measured results
of $P_n(E_{exc})$ as:
\begin{equation}
    P_n(E_{exc}) = 1 - \text{exp}\bigg[\frac{-(E_{exc}-E_{sep})}{\lambda}\bigg],
\end{equation}
with $E_{sep}$ = 7.37 MeV and $\lambda$ = 0.8 MeV. With this correction, and $\omega$ = $E_{exc}$ we get
\begin{equation}
    m(b) = \int d\omega N(\omega, b) \sigma_{abs}(\omega) P_n(\omega) .
\end{equation} 
The corrections on the logarithmic scale are
barely visible, therefore we also show the ratio between $P^c(b)$ and $P^c_{\text{cor}}(b)$
- probability corrected by the mentioned effect,
at the bottom of Fig.\ref{fig:probabilities}. 

\subsection{Neutron emission and $\rho^0$ meson photoproduction}


The $\rho^0$ production in UPC is a process which was measured at RHIC and LHC with large statistics. The cross section for the simultaneous production of $\rho^0$ meson in association with neutron emission can be written as

\begin{eqnarray}
        \sigma_{AA\to AA+\rho+kn+jn}&=& \int \left( \frac{dP_{\gamma  \mathbb{P} }}{dy_\rho} + \frac{dP_{ \mathbb{P} \gamma}}{dy_\rho}  \right)   \nonumber \\
        &\times& S^2(b) P^{c}(b) \, d^2b dy_\rho \;,
\label{eq:photoproduction}
\end{eqnarray}

\noindent where the rapidity-dependent probabilities are:

\begin{equation}
    \frac{dP_{\gamma \mathbb{P} /\mathbb{P}\gamma}}{dy_\rho} = \omega_i N(\omega_i,b) \sigma(\gamma A \to \rho A; W_i) 
\end{equation}
The modeling relay therefore on calculating photoproduction cross section $\sigma(\gamma A \to \rho A; W_i)$ at
phase space dependent $\gamma A$ energies $W_i$ (i = 1,2), $W_i^2=2 \omega_i \sqrt{s_{NN}}$.
In the following, we shall use a simple model as $\rho^0$ production is used only to test our approach for
neutron emissions. In the vector dominance model (VDM):

\begin{equation}
    \sigma(\gamma A \to \rho A; W_i) = \frac{d \sigma_{\gamma A \to \rho A}(W_i; t=0)}{dt} \int^{t_{min}}_{-\infty} |F_A(t)|^2 dt,
\end{equation}

\begin{equation}
    \sigma(\gamma A \to \rho A; W_i)= \frac{\alpha_{em}}{16\pi } \frac{4\pi}{f_\rho^2} \sigma^2_{tot} (\rho A; W_i) .
\end{equation}

\noindent In the classical mechanics Glauber approach, the total cross section for $VA$ collisions reads:
\begin{equation}
    \sigma_{tot} (\rho A; W_i) = \int \left(1-\exp(-\sigma_{tot} (\rho p; W_i) T_A(r))\right) d^2r .
\end{equation}
From the optical theorem one gets:

\begin{equation}
    \sigma_{tot}^2 (\rho p) = 16 \pi \frac{d\sigma (\rho p)}{dt} \bigg|_{t=0} = 16 \pi \frac{f^2_\rho}{4\pi} \frac{1}{\alpha_{em}} \frac{d\sigma (\gamma p \to \rho^0p)}{dt} \bigg|_{t=0},
\end{equation}
We parametrize the forward $\rho^0$ production cross section in elementary collisions as:

\begin{equation}
    \frac{d\sigma (\gamma p \to \rho p)}{dt} \bigg|_{t=0} = B (X W_{\gamma p}^\epsilon + Y W_{\gamma p}^{-\eta}).
    \label{eq14}
\end{equation}

\noindent  For the $\rho^0$ vector meson production we take: $B~=~11$~GeV$^{-2}$, $X=5$~$\mu$b, $\epsilon=0.22$, $Y=26$~$\mu$b, $\eta=1.23$. Here for simplicity we use only one energy-independent slope parameter $B$, identical for Pomeron and
Reggeon exchanges.

So far, extensive investigations have been conducted on the coherent photoproduction of $\rho^0$ vector meson at midrapidity. These experimental studies involved ultraperipehral collisions of Au+Au at RHIC, exploring three distinct center-of-mass energies per nucleon pair: $\sqrt{s_{NN}}=62.4$~GeV \cite{STAR:2011wtm}, $130$~GeV \cite{STAR:2002caw}, and $200$~GeV \cite{STAR:2007elq,STAR:2017enh}. Furthermore, the ALICE experiment at the LHC also examined this reaction in lead-lead UPC at a nucleon-nucleon center-of-mass energy of $\sqrt{s_{NN}}=2.76$~TeV \cite{ALICE:2015nbw} and $5.02$~TeV \cite{ALICE:2020ugp}.

The STAR experiment has derived $d\sigma/dy_V$ cross section by scaling the mutual excitation $(XnXn)$ results due to the poorly known efficiency of the topology trigger. This extrapolation relies on two experimentally measured ratios from the topology sample. The compilation of theoretical results (for two rapidity intervals) with the experimental result is presented in Table~\ref{tab:STAR_ratio}. We start with the ratio of cross sections for different categories for $\sqrt{ s_{NN} }$ = 200 GeV (STAR), see Table~\ref{tab:STAR_ratio}. The first ratio strongly depends on the range of rapidity. In Table~\ref{tab:STAR} we show the corresponding cross section in mb. Our model calculation almost reproduces the experimental data.

\begin{table}
\caption{Ratio of the cross sections for different nuclear neutron configurations, i.e. $\sigma(0n0n)/\sigma(XnXn)$ and $\sigma(0nXn)/\sigma(XnXn)$. Experimental values are from the STAR UPC analysis at $\sqrt{s_{NN}}=200$ GeV, Ref. \cite{STAR:2007elq}.}
\begin{tabular}{l|r|r|r}
\hline
Ratio               & experiment    & theory; $|y_V|<$1   & theory - full $y_V$ \\ \hline
$(0n0n)/(XnXn)$   & $7.1\pm0.3$   & 6.4        &   10.1    \\ 
$(0nXn)/(XnXn)$   & $3.5\pm0.2$   & 3.1       &   3.3           \\  \hline
\end{tabular}
\label{tab:STAR_ratio}
\end{table}

\begin{table}
\caption{Nuclear cross section in $mb$ for the coherent photoproduction of $\rho^0$ in $Au+Au$ UPC at $\sqrt{s_{NN}}=200$ GeV. The experimental values of total cross section $\sigma_{exp}$ are from Ref. \cite{STAR:2007elq}, $|y_V|<1$.}
\begin{tabular}{l|r|r}
\hline
Neutron class   & $\sigma_{exp}$     & $\sigma_{th}$ \\ \hline
$XnXn$        & $14.5 \pm 0.7 \pm 1.9$                & $18.2$              \\
$0n0n$         & $106 \pm 5 \pm 14$\footnote{based on the ratio $(0n0n)/(XnXn)$}                   & $115.6$              \\
\hline
\end{tabular}
\label{tab:STAR}
\end{table}

In Table \ref{tab:STAR2} we show the cross section for the full rapidity range. The experimental data were obtained by
extrapolation from the measured region \cite{STAR:2007elq}. In each of the analyzed kinematical configurations, our simple model systematically overestimates the experimental values. Therefore, a more reasonable approach is to analyze the ratios of the
results.

\begin{table}
\caption{Nuclear cross section in $mb$ for the coherent photoproduction of $\rho^0$ in $Au+Au$ UPC at $\sqrt{s_{NN}}=200$ GeV. The experimental value of total cross section $\sigma_{exp}$ is taken from Ref. \cite{STAR:2007elq}, for the full rapidity range.}
\begin{tabular}{l|r|r}
\hline
Neutron class   & $\sigma_{exp}$     & $\sigma_{th}$ \\ \hline
$XnXn$        & $31.9 \pm 1.5 \pm 4.5$                & $39.3$              \\
$Xn0n+0nXn$         & $105 \pm 5 \pm 15$                    & $130.6$              \\
$0n0n$         & $391 \pm 18 \pm 55$                   & $422.3$              \\
no forward neutron selection
                & $530 \pm 19 \pm 57$                   & $592.1$                \\
\hline
\end{tabular}
\label{tab:STAR2}
\end{table}

Finally in Table \ref{tab:list} we show results for the ALICE experiment \cite{ALICE:2020ugp}
at the LHC with the collision energy $\sqrt{s_{NN}}$ = 5.02 TeV. In contrast to RHIC, the present version of the
used model gives cross sections which are $\sim$25 \% too small. This is caused by too simple parametrization
of~$\frac{d\sigma(\gamma p \to \rho^0 p)}{dt}\bigg|_{t=0}$ (see Eq.~(\ref{eq14})). In the last two columns, we show
experimental and theoretical ratios for a given category defined as:
\begin{equation}
    r_c = \frac{d\sigma^c/dy}{\sum^4_{i=1} d\sigma^i/dy}.
\end{equation}
\noindent The denominator is the sum of contributions from all categories. The "experimental" ratios were obtained from the ALICE data \cite{ALICE:2020ugp}.
The partial probabilities for a given neutron category agree with the experimental data for different
ranges of rapidity.

\begin{table}
\caption{Nuclear cross section $d\sigma/dy$ in $mb$ for the coherent photoproduction of $\rho^0$ in $Pb+Pb$ UPC at $\sqrt{s_{NN}}=5.02$ TeV. The experimental cross section is provided for different rapidity ranges, Ref. \cite{ALICE:2020ugp}.}
\begin{tabular}{l|c|c|c|c}
\hline
Neutron class   & exp.  &  theory & $r^c_{exp}$ [\%]  & $r^c_{theo.}$ [\%] \\ \hline
\multicolumn{5}{c}{$XnXn$}           \\ \hline
$|y|<0.2$           & $24.4 \pm 1.3 ^{+3.4}_{-2.9}$ &  17.92  & 4.54 & 4.39 \\ \hline   
$0.2<|y|<0.45$      & $24.5 \pm 1.2 ^{+3.4}_{-3.0}$ &  21.5  & 4.55  & 4.42 \\ \hline
$0.45<|y|<0.8$      & $25.6 \pm 1.3 ^{+3.5}_{-3.1}$ &  20.5  & 4.68  & 4.49 \\ \hline
\multicolumn{5}{c}{$Xn0n+0nXn$}  \\ \hline
$|y|<0.2$           & $90.2 \pm 1.9 ^{+10.5}_{-9.5}$ & 58.2   &  16.80 & 14.25 \\ \hline   
$0.2<|y|<0.45$      & $87.7 \pm 1.8 ^{+10.2}_{-9.3}$ & 79.6   & 16.28  & 14.31 \\ \hline
$0.45<|y|<0.8$      & $89.9 \pm 2.0 ^{+10.4}_{-9.5}$ & 76.2   & 16.44  & 14.51 \\ \hline
\multicolumn{5}{c}{$0n0n$}            \\ \hline
$|y|<0.2$           & $431.1 \pm 4.0 ^{+36.8}_{-33.6}$ & 332.3   &  80.28 & 81.39 \\ \hline   
$0.2<|y|<0.45$      & $433.8 \pm 3.8 ^{+37.0}_{-33.8}$ & 395.1   &  80.54 & 81.25 \\ \hline
$0.45<|y|<0.8$      & $436.7 \pm 4.2 ^{+37.3}_{-34.0}$ & 369.5   &  79.84 & 80.98 \\ \hline
\multicolumn{5}{c}{no forward neutron selection}  \\ \hline
$|y|<0.2$           & $537.0 \pm 4.6 ^{+46.1}_{-42.0}$  &  408.3 &  &  \\ \hline   
$0.2<|y|<0.45$      & $538.6 \pm 4.4 ^{+46.2}_{-42.1}$  &  486.3 &  &  \\ \hline
$0.45<|y|<0.8$      & $547.0 \pm 4.9 ^{+46.9}_{-42.8}$  &  456.3 &  &  \\ \hline
\hline
\end{tabular}
\label{tab:list}
\end{table}


Having shown that our approach correctly reproduces fractions of
cross sections for a particular neutron category, we can proceed to the case of interest, that is emission of neutrons associated
with the $\gamma\gamma \to \gamma\gamma$ scattering.

\subsection{Neutron emission and light by light scattering}

Compared to the experimental and theoretical analysis of vector meson photoproduction
associated with neutron emission, such an analysis was not done for LbL scattering.
We start by calculating the cross section for different neutron classes and
photons in the ATLAS acceptance region: $y_1, y_2 \in$ (-2.4,2.4) and
$p_{t,\gamma} >$ 2.5 GeV at the future collision energy $\sqrt{s_{NN}}$ = 5.36 TeV. 
We use the same model, which was discussed in detail, Ref. \cite{Jucha2024}.
and gives relatively good agreement with the experiment. Moving to 
the main issue of the study, in Tab. \ref{tab1} we show a summary of the
ratios for the given neutron categories. The cross section for the $0n0n$ category is
the biggest and constitutes 75 \% of the total fiducial cross section.
The cross section for one-side neutron emission constitutes 9.75 \% + 9.75 \%
= 19.5 \% and the cross section for both side emission is only 5.2 \%.
These numbers can be compared with similar fractions of neutron classes for $\rho^0$
production in UPC (see the previous subsection). In general, the fractions
in Tables \ref{tab:list} and \ref{tab1} are similar but not identical.

\begin{table}
\centering
\caption{Cross section in nb for light by light scattering in $Pb+Pb$ UPC and different neutron categories. Here $\sqrt{s_{NN}} = 5.36$ TeV corresponding to the ongoing ATLAS analysis.}
\label{tab1}
\begin{tabular}{c|c c c c }
\hline
              &  $\sigma_{total}$         & $0n0n$ & $XnXn$ & $Xn0n+0nXn$ \\ \hline
cross section [nb]  & 81.886     & 61.679 & 4.261 & 15.963  \\
percent of $\sigma_{total}$ [\%]     &        &  75.30  & 5.20 &  9.50  \\ \hline
\end{tabular}
\end{table}


\begin{figure}[!h]
    \centering
    \includegraphics[width=0.46\textwidth]{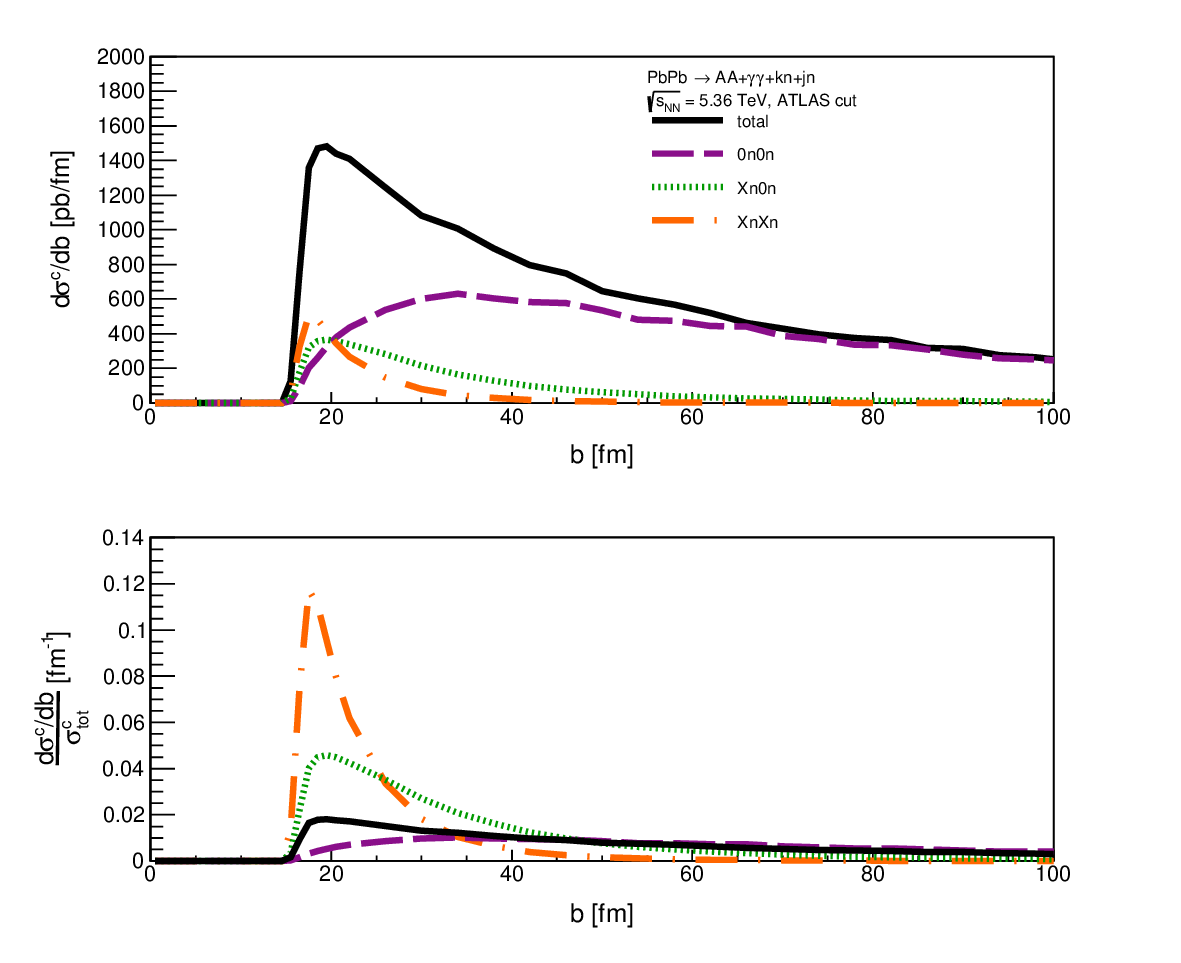}
    \caption{Impact parameter distribution for: total
      cross section (solid line), $0n0n$ (dashed line), $Xn0n$ (dotted
      line) and $XnXn$ (dash-dotted line). }
      \label{fig:dsig_db}
\end{figure}


\begin{figure}[!h]
    \centering
    \includegraphics[width=0.42\textwidth]{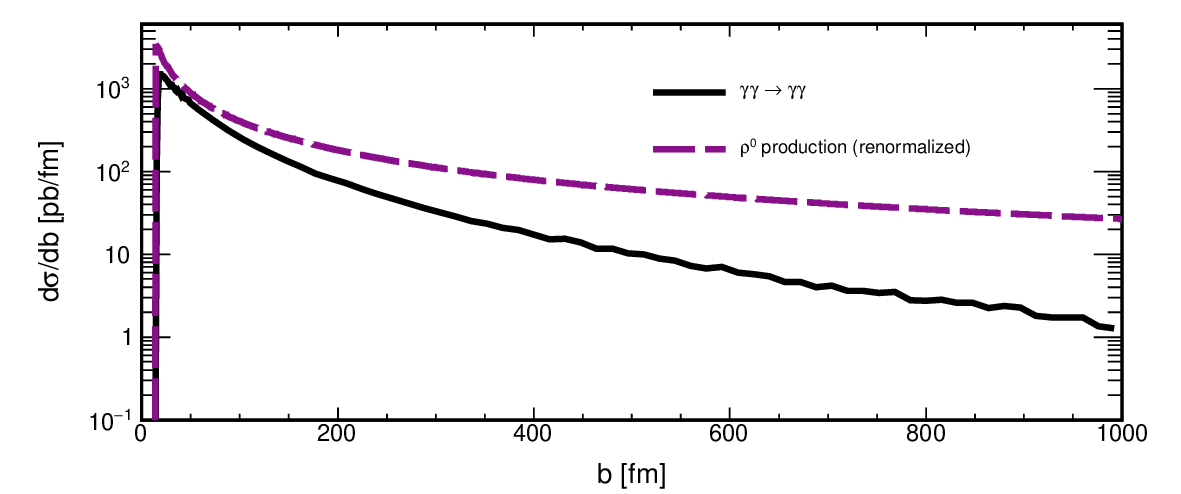}
    \caption{Impact parameter distribution for $\rho^0$ production ($\sqrt{s_{NN}}$ = 5.02 TeV, $|y| < 1$) and
    for $\gamma\gamma \to \gamma\gamma$ ($\sqrt{s_{NN} }$ = 5.36 TeV, $|y| <$ 2.4, $p_t >$ 2.5 GeV). The distribution for $\rho^0$ production is arbitrarily renormalized.}
      \label{fig:dsig_db_comp}
\end{figure}


The numbers in Table \ref{tab1} can be better understood by inspecting
Fig. \ref{fig:dsig_db} where the dependence of the cross section on impact parameter is shown.

\begin{figure}[!h]
    \centering
    \includegraphics[width=0.46\textwidth]{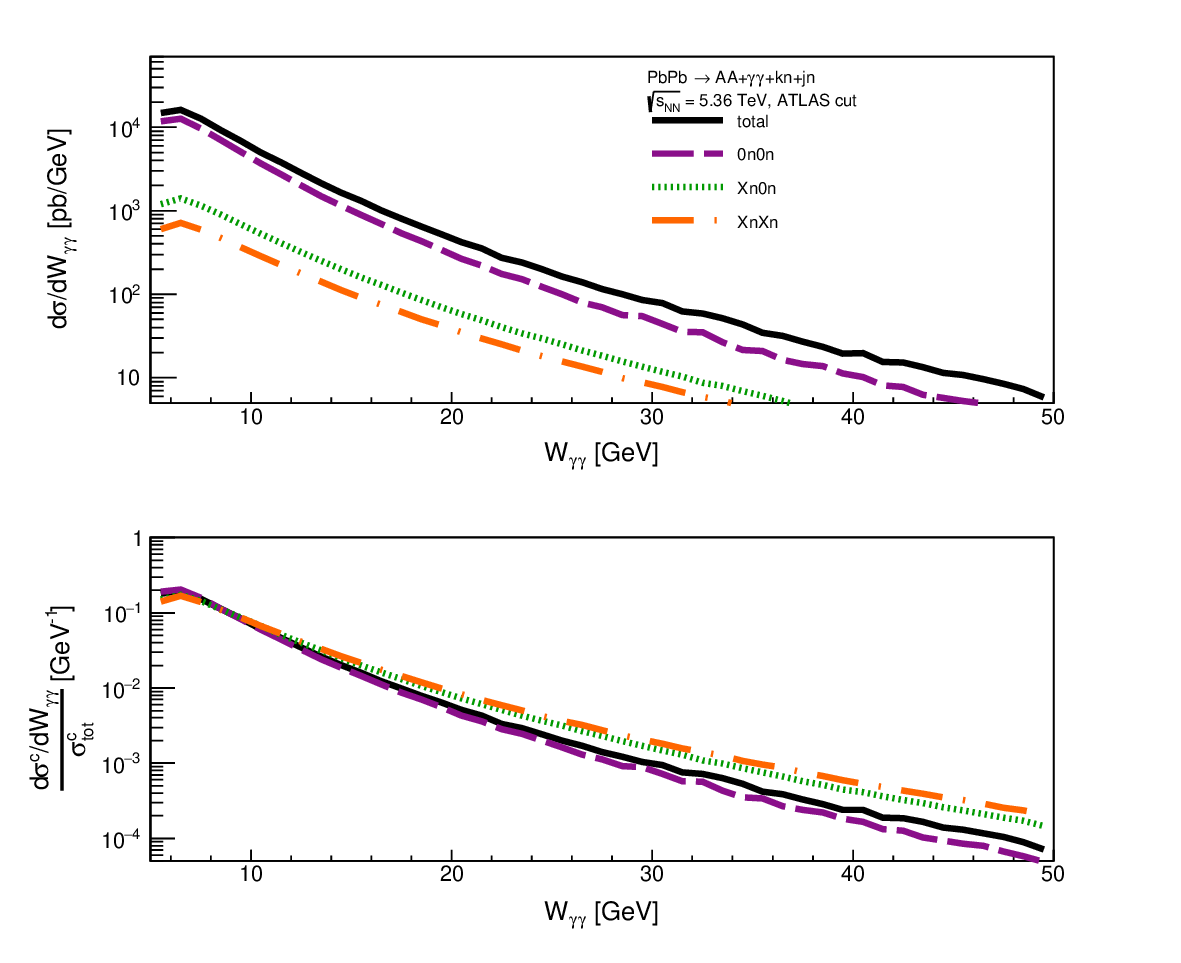}
    \caption{Diphoton mass distribution of cross section for: total
      cross section (solid line), and $0n0n$ (dashed line), $Xn0n$ (dotted
      line) and $XnXn$ (dash-dotted line) categories. }
      \label{fig:dsig_dM}
\end{figure}

\begin{figure}[!h]
    \centering
    \includegraphics[width=0.46\textwidth]{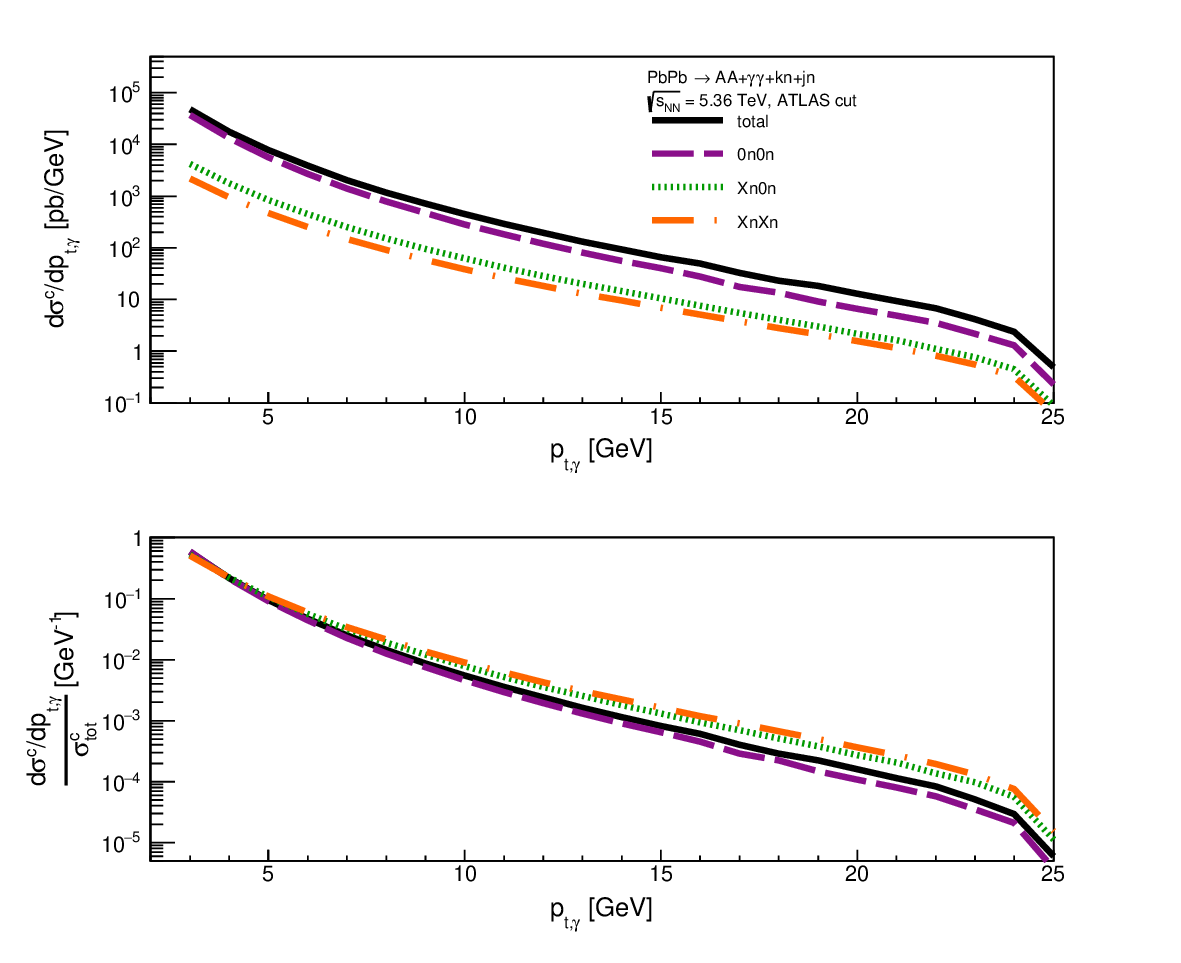}
    \caption{Transverse momentum distribution for:
      total cross section (solid line), and $0n0n$ (dashed line), $Xn0n$
      (dotted line) and $XnXn$ (dash-dotted line) categories. }
      \label{fig:dsig_dpt}
\end{figure}

We show the cross section for different categories.
While the $0n0n$ category is related to the broad range of impact
parameter, the emission from both nuclei happens at small impact
parameters, close to the surface of nuclei ($b\sim2R_A$). The fluctuations visible in
the figure are due to the Monte Carlo method used in our calculation. The distribution in $b$ is rather
theoretical and cannot be measured. The percentage of $0n0n$, $Xn0n+0nXn$ categories here is smaller than for
$\rho^0$ production (see Table \ref{tab:list}). This is associated with
relatively larger impact parameters in the $\rho^0$ production than for
the $\gamma\gamma \to \gamma\gamma$ scattering (see Fig. \ref{fig:dsig_db_comp}).
The distribution for $\rho^0$ was renormalized arbitrarily in order
to compare shapes of distributions.
\begin{figure}[H]
    \centering
    \includegraphics[width=0.46\textwidth]{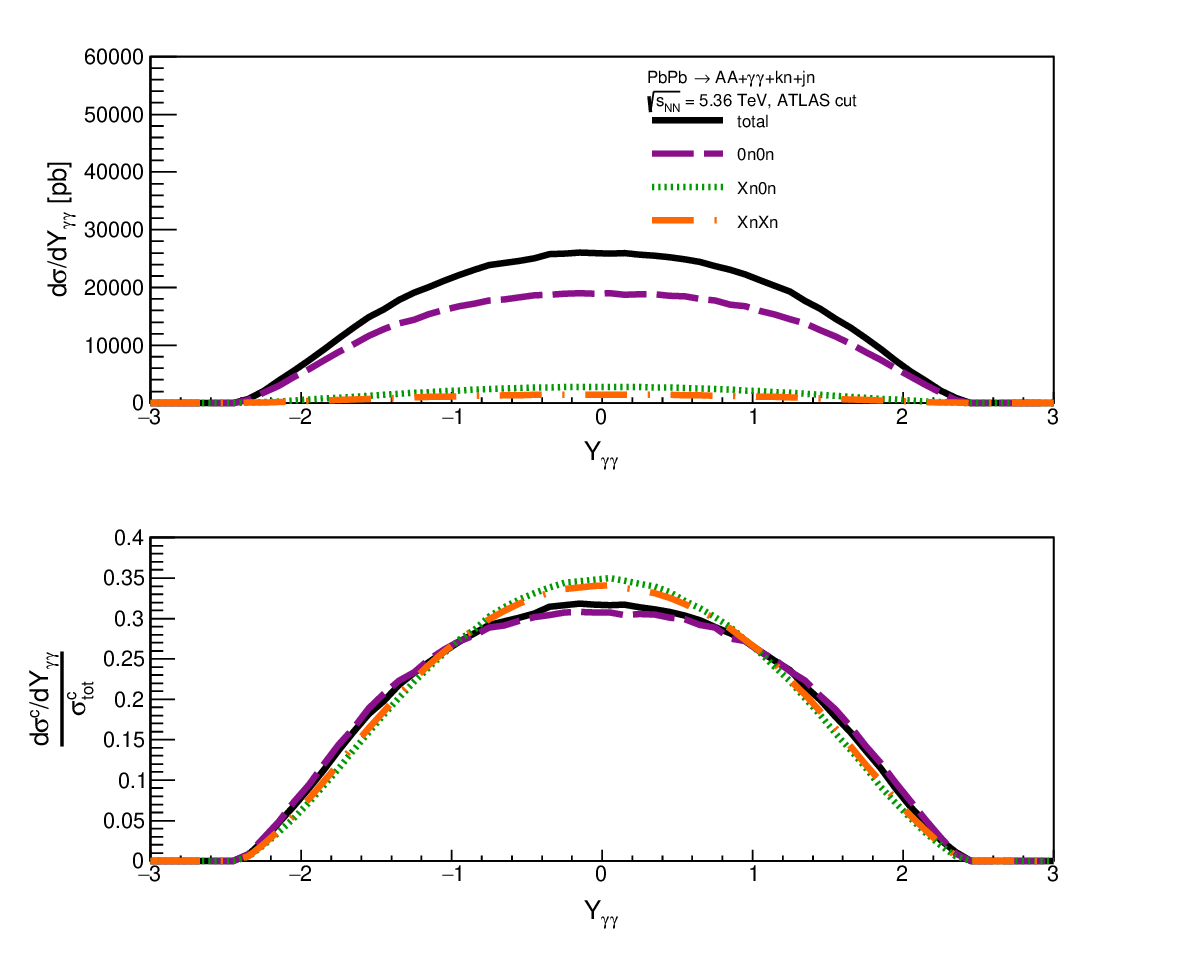}
    \caption{Rapidity distribution of cross section for: total cross
      section (solid line), and $0n0n$ (dashed line), $Xn0n$ (dotted line) 
      and $XnXn$ (dash-dotted line) categories. }
      \label{fig:dsig_dy}
\end{figure}

\begin{figure}[H]
    \centering
    \includegraphics[width=0.46\textwidth]{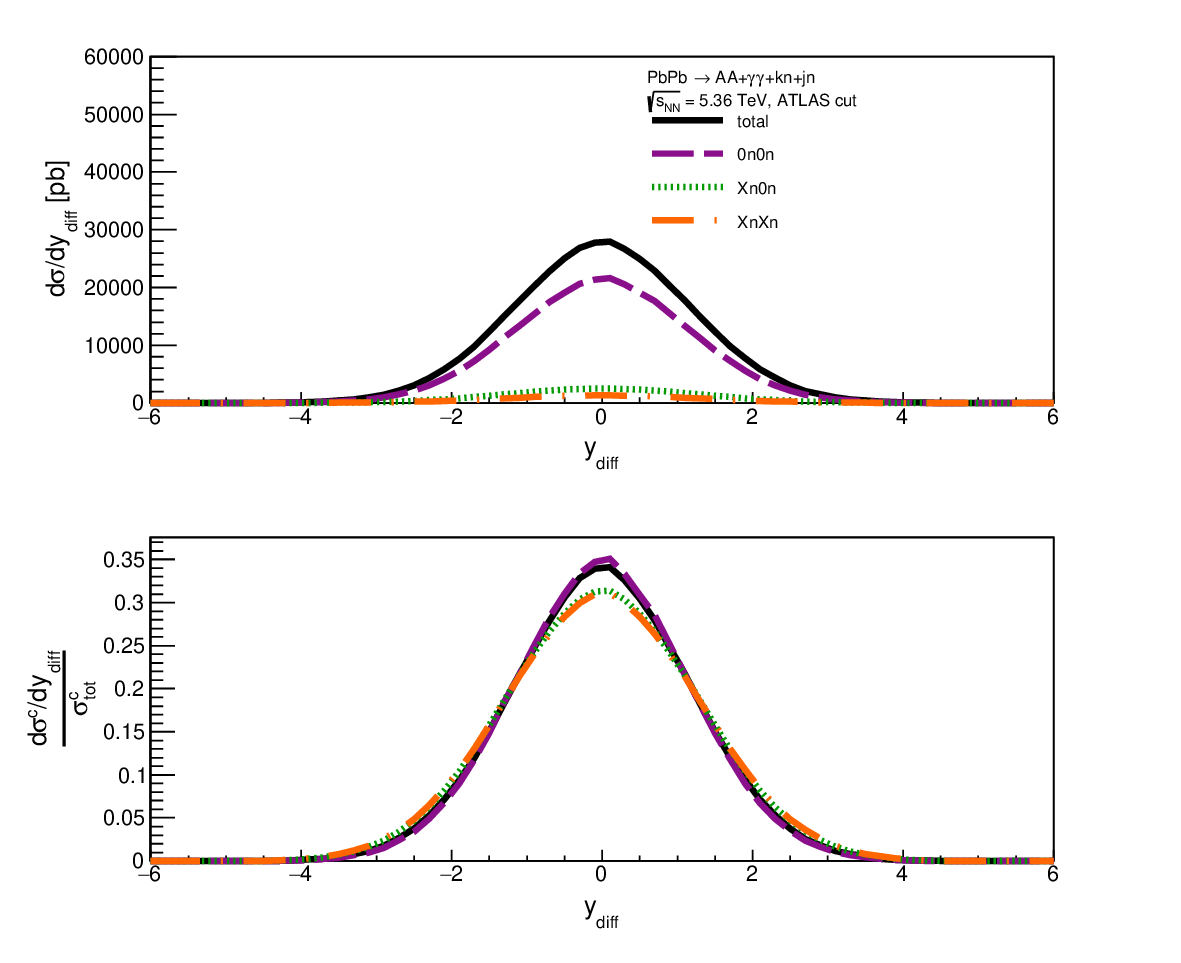}
    \caption{Rapidity difference distribution of cross section for:
      total cross section (solid line), and $0n0n$ (dashed line), $Xn0n$
      (dotted line) and $XnXn$ (dash-dotted line) categories. }
    \label{fig:dsig_dydiff}
\end{figure}

For example, in the following we shall show distributions in possible to measure
observables. In Fig.\ref{fig:dsig_dM} we show the distribution in diphoton
invariant mass. Taking into consideration only relative contribution,
the shape of the distribution weakly depends on the configuration under
consideration. Naturally, the largest contribution to the 
total cross section comes from the situation when no neutron is emitted
from any nucleus.

In Fig.\ref{fig:dsig_dpt}, Fig.\ref{fig:dsig_dy},
Fig.\ref{fig:dsig_dydiff} we show respectively distributions in photon
transverse momentum, diphoton rapidity and distribution in 
$y_{diff} = y_1 - y_2$. In all cases in the upper panels, we show
absolutely normalized cross sections and in the lower panels the
distributions normalized to unity in order to compare the shape of the
distributions for different neutron classes. By careful inspection of
the lower panels we see that the shapes of different observables
depend on categorization of neutron emissions.

\section{Summary and Conclusions}

In the present paper we have discussed in detail the production of
neutrons associated with $\gamma \gamma \to \gamma \gamma$
scattering in UPC of $^{208}Pb + ^{208}Pb$. We have presented
the cross sections for different neutron categories and for photon cuts of the respective recent ATLAS experiment.
We have also shown fractions of cross sections for different
neutron categories and compared them with similar fractions 
for $\rho^0$ meson production in UPC. In general,
the fractions for both cases ($\rho^0$ production and $\gamma\gamma \to \gamma\gamma$)
are similar, but not identical.

We have presented the impact parameter dependence of
the probability of a given neutron category ($0n0n$, $0nXn+Xn0n$, $XnXn$).
We have also briefly discussed the impact of a recent measurement
of the ESR collaboration at GSI/FAIR Darmstadt of the
probability of neutron emissions at low excitation energies close
to neutron separation energy.
We have found that there are rather small corrections to the cross 
section at the \% level.

We have calculated several differential distributions
in photon variables (rapidity, transverse momentum, diphoton
invariant mass and rapidity difference) separately for different
neutron categories. We have demonstrated how the shape of the 
distribution depends on the choice of neutron categories.

We hope our predictions will be confronted by the ATLAS
experimental results in the near future.
In the present paper, we have considered the dominant coherent 
contribution, called sometimes elastic-elastic or double elastic.
The potential contribution of inelastic processes 
(photon couples to nucleons) discussed by us recently \cite{KGS_2025} 
requires a special attention which goes beyond the scope of the present letter. 
The deviation of experimental numbers from our predictions
on fractions of topological cross sections may signal a participation of 
inelastic contributions \cite{KGS_2025}.
We suggest that the ATLAS collaboration could measure the fractions of
cross sections also for larger transverse momenta of diphotons $p_{t,\gamma\gamma}$
or larger acoplanarities than used for selection of the $\gamma\gamma\to\gamma\gamma$ scattering.
This would provide interesting information on the background contribution which is not well
understood so far in our opinion.

\vspace{0.5cm}

{\bf Acknowledgment}

We are indebted to Michał Ciemała for pointing to us the
recent GSI result \cite{GSI} on the probability of the neutron emission at low excitation energies
and Iwona Grabowska-Bołd for a discussion of the ATLAS experiment
and plans of the ATLAS collaborations for the future.

\bibliographystyle{unsrtnat}
\bibliography{biblio}

\end{document}